\documentclass[
  prl,
  onecolumn,
  showpacs,
  preprintnumbers,
  amsmath,
  amssymb,
  nofootinbib,
  floatfix
]{revtex4-1}
\usepackage{graphicx}
\usepackage{mathrsfs}
\usepackage{hyperref}

\usepackage{slashed}

\usepackage{color}

\usepackage{gensymb}

\usepackage{amsmath}
\allowdisplaybreaks[4]

\def \matrix #1 {\left(\begin{array}{cc} #1 \end{array}\right)}

\def\II{\hbox{{1}\kern-.25em\hbox{l}}}

\begin{document}

\title{Next-to-Leading-Order QCD Predictions for the $\Sigma$ Dirac Form Factors}

\author{Bo-Xuan Shi}
\email{corresponding author: shibx@mail.nankai.edu.cn}

\author{Hui-Xin Yu}
\email{corresponding author: yuhuixin@mail.nankai.edu.cn}

\author{Xue-Chen Zhao}
\email{corresponding author: zxc@mail.nankai.edu.cn}

\affiliation{\vspace{0.2 cm}
School of Physics, Nankai University, \\
Weijin Road 94, Tianjin 300071, P.R. China \\}

\date{\today}

\begin{abstract}
In this work, we compute the next-to-leading-order QCD corrections to the Dirac electromagnetic form factors of the $\Sigma$ hyperons within the hard-collinear factorization framework at leading power.
The corresponding short-distance coefficient functions are extracted from the relevant seven-point partonic correlation functions.
We find that the one-loop radiative corrections to the leading-twist hard-scattering contributions are numerically significant over a broad range of momentum transfer.
Combining the perturbatively calculated hard kernels with nonperturbative $\Sigma$ distribution amplitudes determined from lattice QCD, we present state-of-the-art theoretical predictions for the $\Sigma$ hyperon electromagnetic form factors.

\end{abstract}

\maketitle

%
\section{Introduction}
The hyperon electromagnetic form factors characterize the internal electromagnetic structures of the hyperon, which are scalar functions of the momentum transfer $Q^2$. The systematic study of the hyperon electromagnetic form factors is important for deepening our understanding of the hard-collinear factorization framework. It is also crucial for achieving a deep understanding of nonperturbative QCD effects that govern quark confinement in baryons.
The tree-level hard-scattering kernel of baryon electromagnetic form factors within the hard-collinear factorization approach has been known since the early developments of perturbative QCD nearly half a century ago~\cite{Lepage:1979za,Chernyak:1984bm}.
Since then, several re-computations of these results have been carried out~\cite{Ji:1986uh,Carlson:1987sw,Stefanis:1987vr,Brooks:2000nb,Thomson:2006ny}. 
The underlying model-independent perturbative factorization framework is based on diagrammatic analyses~\cite{Lepage:1980fj,Efremov:1979qk,Chernyak:1983ej} and relevant effective field theories~\cite{Kivel:2010ns,Kivel:2012mf}.
From the experimental perspective, there has also been considerable interest in assessing the reliability of perturbative QCD predictions for this process~\cite{Isgur:1988iw,Radyushkin:1998rt,Bolz:1996sw,Brodsky:1989pv,Botts:1989kf,Li:1992nu}.

Recently, the next-to-leading-order (NLO) QCD corrections to the nucleon electromagnetic form factors and NNLO QCD corrections to the pion electromagnetic form factors have been computed along the same lines~\cite{Huang:2024ugd, Ji:2024iak}, employing a systematic prescription for the proper treatment of evanescent operators~\cite{Dugan:1990df,Herrlich:1994kh,Wang:2017ijn,Gao:2021iqq}, as well as within the so-called KM scheme~\cite{Krankl:2011gch,Gracey:2012gx} for the nucleon electromagnetic form factors.
Independent calculations of the NLO corrections to the nucleon electromagnetic form factors have also been carried out within the KM scheme~\cite{Chen:2024fhj}.
Very recently, the two-loop renormalization-group (RG) evolution of the nucleon distribution amplitude has been determined for the first time~\cite{Huang:2025bbk}, and the corresponding result is equally applicable to hyperons.
This result provides the last missing ingredient for a complete next-to-leading-logarithmic (NLL) analysis of nucleon form factors within the hard-collinear factorization framework. 
When combined with the NLO hard-gluon-scattering analysis carried out in this work, it also enables a consistent NLL resummation for hyperon electromagnetic form factors.

Experimentally, there are many effects in measuring the baryon electromagnetic form factors in the spacelike region
~\cite{Gao:2003ag, Hyde:2004gef, Arrington:2006zm, Pacetti:2014jai, Gross:2022hyw, JeffersonLabHallA:1999epl}
. Owing to the short lifetimes of hyperons, it is difficult to prepare hyperon targets, which makes direct measurements of spacelike form factors highly challenging. However, in the timelike region, the electromagnetic form factors can be accessed through electron–positron annihilation processes, which have been measured by the BESIII and Belle collaborations~\cite{BESIII:2017hyw,BESIII:2019cuv,BESIII:2020ktn,BESIII:2021aer,BESIII:2020uqk,BESIII:2021rkn,Wang:2022zyc,Belle:2022dvb}.

In this work, we apply the same techniques in~\cite{Huang:2024ugd} to investigate the $\Sigma$ electromagnetic form factors within the hard-collinear factorization formalism. We focus on the spacelike form factors, while the timelike counterparts can be obtained via appropriate replacement of the momentum transfer $Q^2 \to -Q^2$.
As the timelike form factors involve additional contributions from intermediate hadronic states, we limit the present study to a theoretical numerical analysis of the spacelike electromagnetic form factors.

We first employ the modern QCD factorization formalism to analytically extract the NLO short-distance coefficient functions relevant for the hard-scattering contributions to hyperon electromagnetic form factors. The calculation is carried out by evaluating the required seven-point partonic amplitudes at $\mathcal{O}(\alpha_s^3)$, making use of state-of-the-art one-loop computational techniques that are by now standard in higher-order perturbative QCD analyses.
A careful implementation of ultraviolet (UV) renormalization and infrared (IR) subtractions is then performed, ensuring a rigorous treatment that is fully consistent with the underlying factorization framework. This allows us to isolate the perturbatively calculable hard-scattering kernels in a transparent manner.
Based on the analytically determined NLO hard coefficient functions, we subsequently investigate the phenomenological impact of the corresponding radiative corrections on the hard-gluon-exchange contributions to hyperon electromagnetic form factors at large momentum transfers. 
In particular, we present numerical predictions obtained with a representative model of the leading-twist hyperon distribution amplitude, thereby assessing the size of the NLO effects.

%
\section{Electromagnetic Hyperon Form Factors}
%
Adopting the customary definitions of the hyperon electromagnetic form factors allows us to parametrize the hyperon matrix element of the electromagnetic current in terms of scalar form factors~\cite{Foldy:1952zz}
\begin{align}
&\langle \Sigma(p^{\prime})  | J_{\mu} | \Sigma(p) \rangle = \bar \Sigma(p^{\prime}) \,  \big [ \gamma_{\mu} \, F_1 (Q^2) - \, i \, \frac{\sigma_{\mu \nu} \, q^{\nu}}{m_{\Sigma}}  \, F_2 (Q^2) \big  ]  \, \Sigma(p) \,,
\label{definition of nucleon form factors}
\end{align}
with the electromagnetic current given by
$J_{\mu}  =  e_{u} \, \bar u(x) \gamma_{\mu} u(x) +  e_{d} \, \bar d(x) \gamma_{\mu} d(x) +  e_{s} \, \bar s(x) \gamma_{\mu} s(x)$, where we have closed the irrelevant quark flavors.
The $q = p - p^{\prime}$ denotes the momentum transfer carried by the virtual photon,  while the $p$ and $p^{\prime}$ denote the four-momenta of the initial and final hyperon states, respectively. The $m_\Sigma$ is the mass of the $\Sigma$ hyperon. We also define $Q^2 = -q^2$, which is positive in the spacelike region.

In the large momentum transfer region, the momenta of the initial- and final-state hyperons $p$ and $p'$, are nearly light-like and thus lie close to the light-cone.
It is therefore convenient to introduce two light-cone vectors $n$ and $\bar n$, which satisfy $n \cdot n = 0$, $\bar n \cdot \bar n = 0$, and $n \cdot \bar n = 2$,
in order to decompose the hyperon momenta $p$ and $p'$. Explicitly, in the limit $Q^2 \to \infty$, one has $p_{\mu} =  ({n \cdot p} /2) \, \bar n_{\mu}$
and $p^{\prime}_{\mu} = ({\bar n \cdot p^{\prime}} /2) \, n_{\mu}$.

At the large momentum transfer, the hard-gluon-exchange contribution to the helicity-conserving Dirac form factor exhibits the asymptotic scaling behavior $1/Q^4$.
However, the helicity-flip Pauli form factor exhibits the asymptotic scaling behavior $1/Q^6$, which is suppressed compared with the Dirac form factor.
As a result, the Dirac hyperon form factor $F_1(Q^2)$ is of leading power, and we will only consider the hard-gluon-exchange contribution to the $F_1 (Q^2)$ in the following analysis.

\section{Factorization Formalism}
Within the hard-collinear factorization formalism~\cite{Lepage:1980fj,Chernyak:1983ej},
The leading power contribution to the Dirac form factor at large momentum transfer
admits the following representation
\begin{align}
Q^4 F_1(Q^2) =& {(4 \, \pi \alpha_s)^2} \iint  {\cal D} x  \, {\cal D}  y \Bigg [
 \varphi_\Sigma(x_i, \mu_F) \, T_{\Sigma}(x_i,  y_i,  Q^2, \mu_F)  \, \varphi_\Sigma(y_i, \mu_F) + 
 \varphi_T(x_i, \mu_F) \, T_T(x_i,  y_i,  Q^2, \mu_F)  \, \varphi_T(y_i, \mu_F) \Bigg ]
\nonumber\\
=& {(4 \, \pi \alpha_s)^2} \iint  {\cal D} x  \, {\cal D}  y \Bigg [
 \varphi_\Sigma(x_i, \mu_F) \, H_{\Sigma} (x_i,  y_i,  Q^2, \mu_F) \,  \varphi_\Sigma(y_i, \mu_F) \Bigg ] \,,
\label{hard-collinear factorization formula}
\end{align}
where the integration measure is defined as $
{\cal D} x  = d x_1 \, d x_2 \, d x_3  \,\,   \delta(\sum_i x_i - 1)$.
The $\mu_F$ denotes the factorization scale associated with the resolution at which the hyperon structure is probed.
We have made use of isospin symmetry~\cite{Chernyak:1984bm,King:1986wi,Chernyak:1987nv} in \eqref{hard-collinear factorization formula} , which imposes an additional constraint
\begin{align}
    &\varphi_T(x_1, x_2, x_3, \mu_F) = \frac{1}{2}\,\Big[
    \varphi_\Sigma(x_1, x_3, x_2, \mu_F) + \varphi_\Sigma(x_2, x_3, x_1, \mu_F) \Big] \,,
    \label{isospion}
\end{align}
so that the three hard-scattering kernels are related by
\begin{align}
H_{\Sigma}(x_1, x_2, x_3,  y_1, y_2, y_3,  Q^2, \mu_F)
=& {1 \over 4} \, \bigg [ T_{T}(x_1, x_3, x_2,  y_1, y_3, y_2,  Q^2, \mu_F) + T_{T}(x_3, x_1, x_2,  y_3, y_1, y_2,  Q^2, \mu_F)
\nonumber\\
&+  T_{T}(x_3, x_1, x_2,  y_1, y_3, y_2,  Q^2, \mu_F) + T_{T}(x_1, x_3, x_2,  y_3, y_1, y_2,  Q^2, \mu_F)  \bigg ]
\nonumber\\
&+ T_{\Sigma}(x_1, x_2, x_3,  y_1, y_2, y_3,  Q^2, \mu_F)
.
\end{align}

The factorization formula~\eqref{hard-collinear factorization formula} shows that the form factor $F_1(Q^2)$ can be factorized into the convolution of the short-distance coefficient and the $\Sigma$ distribution amplitudes. The short-distance coefficient function $H_\Sigma$ can be calculated perturbatively
$
H_{\Sigma} = \sum_{\ell=0}^{\infty} \left ({\alpha_s \over 4 \, \pi} \right )^{(\ell)} H_{\Sigma}^{(\ell)}
$.
The perturbative expansion in $\alpha_s$ for the other quantities, such as $T_\Sigma$ and $T_T$, follows the same convention as for $H_\Sigma$.
The twist-three hyperon distribution amplitudes $\varphi_\Sigma$ and $\varphi_T$ are typical nonperturbative quantities, which can be defined by the renormalized matrix elements of the three-body collinear operators~\cite{Chernyak:1983ej,Braun:2000kw}
\begin{align}
& \left \langle 0 \left |   \epsilon_{i j k} \left [  u_{i}^{\uparrow}(t_1  n) C \slashed{n} u_{j}^{\downarrow}(t_2 n)  \right ] 
\slashed{n}  s_{k}^{\uparrow}(t_3 n)
\right | \Sigma^{\uparrow} (p) \right  \rangle =  - {f_\Sigma(\mu_F) \over 2} (n \cdot p) \slashed{n} \Sigma^{\uparrow}(p)
\int {\cal D} x {\rm exp} \left [ - i  n \cdot p  \sum_{i=1}^{3} x_i t_i \right ]\, \varphi_\Sigma(x_i, \mu_F) \,,
\nonumber\\
&\left \langle 0 \left | \epsilon_{i j k} \left [  u_{i}^{\uparrow}(t_1 n)  C \gamma_{\perp \alpha} \slashed{n}
u_{j}^{\downarrow}(t_2 n) \right ]  \gamma_{\perp}^{\alpha}
\slashed{n}  s_{k}^{\uparrow}(t_3 n) \right | \Sigma^{\uparrow} (p) \right  \rangle = 2 f_\Sigma(\mu_F)  (n \cdot p)  \slashed{n} \Sigma^{\uparrow}(p)
\int {\cal D} x  {\rm exp} \left [ -  i  n \cdot p   \sum_{i=1}^{3} x_i t_i \right ] \varphi_T(x_i, \mu_F)  \,,
\label{definiton of the twist-3 nucleon DA}
\end{align}
where $C$ denotes the charge-conjugation matrix, %
and the ``$\uparrow$" and ``$\downarrow$" indicate the chirality of the quark fields, explicitly, $q^{\uparrow} = \frac{1}{2}(1 + \gamma_5)\, q$ and $q^{\downarrow} = \frac{1}{2}(1 - \gamma_5)\, q$.
In addition, the three finite-length Wilson lines ensuring gauge invariance~\cite{Braun:1999te} are omitted for brevity.

The RG equation governing the physical operators appearing on the left-hand side of~\eqref{definiton of the twist-3 nucleon DA}~\cite{Balitsky:1987bk} provides the motivation for performing a conformal expansion of $\varphi_\Sigma$ in terms of a complete set of orthogonal polynomials
\begin{align}
\varphi_\Sigma(x_i, \mu_F)  = 120 \, x_1 x_2 x_3  \, \sum_{n=0}^{\infty} \, \sum_{k=0}^{n} \,
\varphi_{n k}(\mu_F)  \, {\cal P}_{nk}(x_i) \,,
\label{eq:conformal expansion}
\end{align}
These polynomials $\mathcal{P}_{nk}$ are defined as eigenfunctions of the corresponding one-loop evolution kernel. Our choice of the first six polynomials ${\cal P}_{nk}$ are~\cite{Braun:2008ia}
\begin{align}
    & \mathcal P_{00} = 1 \, ,\, && \mathcal P_{10} = 21(x_1 - x_3) \, ,
\nonumber \\
    & \mathcal P_{11} = 7(x_1 - 2 x_2 + x_3)\,  ,\, && \mathcal P_{20} = {63 \over 10} \left [ 3(x_1 - x_3)^2 - 3 x_2(x_1 + x_3) + 2 x_2^2 \right ] \, ,
\nonumber \\
    & \mathcal P_{21} = {63 \over 2} \left ( x_1 - 3 x_2 + x_3 \right ) \left ( x_1 - x_3 \right )\,  ,  \,  &&	 \mathcal P_{22} = {9 \over 5} \left [ x_1^2 + 9 x_2(x_1 + x_3) -12 x_1 x_3 - 6 x_2^2 + x_3^2 \right ] \, .
    \label{pnk}
\end{align}
The polynomials ${\cal P}_{nk}$ exhibit a definite parity symmetry under the interchange $x_1 \leftrightarrow x_3$, which can be readily verified from~\eqref{pnk}.
Till now, we have established the theoretical framework for constructing the QCD factorization formulae for hyperon electromagnetic form factors at large momentum transfers.

\section{Extraction of short-distance coefficients}
The short-distance matching coefficient $H_\Sigma$ may be obtained in a straightforward manner by evaluating the following seven-point QCD matrix element in $D$ dimensions
\begin{align}
\Pi_{\mu} =&  \langle u(p_1^{\prime}) \, u(p_2^{\prime}) \, s(p_3^{\prime})
| J_{\mu} |  u(p_1) \, u(p_2) \, s(p_3) \rangle \,.
\label{corr}
\end{align}
At leading power accuracy, the external parton momenta are restricted to their dominant light-cone components, i.e.
$p_i = x_i \, p$ and $p_i^{\prime} = y_i \, p^{\prime}$.
We employ dimensional regularization with $D = 4 - 2\,\epsilon$ to regularize both UV and IR divergences.
The UV divergences are removed by renormalizing the strong coupling $\alpha_s$ in the ${\rm \overline{MS}}$ scheme
\cite{Beneke:2008ei},
whereas the IR singularities are subtracted by the hyperon distribution amplitude contributions, with the aid of a consistent prescription for the treatment of evanescent operators
\cite{Dugan:1990df,Herrlich:1994kh,Wang:2017ijn,Gao:2021iqq}.

The computation of the correlation function~\eqref{corr} is the same as the case of nucleon~\cite{Huang:2024ugd}, which suggests a complete operator basis of the form
\begin{align}
{\cal O}_{\Sigma}^{\mu}  =& [ \bar \chi_{u}^{\downarrow}   \slashed{\bar n}  C^{-1} \bar \chi_{u}^{\uparrow}  ]  \,
 [ \bar \chi_{s}^{\uparrow}  \slashed{\bar n} \,  \gamma_{\perp}^{\mu} \,
 \slashed{n}   \xi_{s}^{\uparrow}  ] \,
 [  \xi_{u}^{\uparrow}  C \slashed{n}  \xi_{u}^{\downarrow} ],
\nonumber \\
{\cal O}_{T}^{\mu}  =&  [ \bar \chi_{u}^{\uparrow}  \slashed{\bar n}  \gamma_{\perp \beta}  C^{-1}
\bar \chi_{u}^{\uparrow}  ] \, [ \bar \chi_{s}^{\downarrow} \, \slashed{\bar n}   \gamma_{\perp}^{\beta}  \, \gamma_{\perp}^{\mu}  \, \gamma_{\perp}^{\alpha}  \slashed{n}  \xi_{s}^{\downarrow}] \, [ \xi_{u}^{\uparrow} C  \gamma_{\perp \alpha} \slashed{n} \xi_{u}^{\uparrow} ],
\nonumber \\
E_{1}^{\mu}  =& [\bar \chi_u \gamma_{\perp}^{\mu} \xi_u  ] \, [\bar \chi_u \gamma_{\perp}^{\alpha} \xi_u ] \, [\bar \chi_s \gamma_{\perp \alpha} \xi_s ] - {1 \over 8} \, {\cal O}_{\Sigma}^{\mu} -  {1 \over 2} \, {\cal O}_{T}^{\mu},
\nonumber \\
E_{2}^{\mu}  =&  [\bar \chi_u \gamma_{\perp}^{\alpha} \gamma_{\perp}^{\beta} \gamma_{\perp}^{\mu} \xi_u  ] \, [\bar \chi_u \gamma_{\perp \beta}  \xi_u ] \, [\bar \chi_s \gamma_{\perp \alpha}  \xi_s ]   -  {\cal O}_{T}^{\mu},
\nonumber \\
E_{3}^{\mu}  =&   [\bar \chi_u \gamma_{\perp \alpha}  \xi_u  ] \, [\bar \chi_u \gamma_{\perp \beta}  \xi_u ] \, [\bar \chi_s \gamma_{\perp}^{\beta}  \gamma_{\perp}^{\alpha} \gamma_{\perp}^{\mu}  \xi_s ] - {1 \over 4} \, {\cal O}_{\Sigma}^{\mu},
\nonumber\\
F_{1}^{\mu}  =&  [ \bar \chi_{u}^{\downarrow}  \slashed{\bar n}  C^{-1} \bar \chi_{u}^{\uparrow}  ] \, [ \bar \chi_{s}^{\uparrow}  \slashed{\bar n} \,  \gamma_{\perp}^{\mu} \, \gamma_{\perp}^{\alpha} \gamma_{\perp}^{\beta} \, \slashed{n}   \xi_{s}^{\uparrow}  ] \, [  \xi_{u}^{\uparrow}  C \gamma_{\perp \beta} \gamma_{\perp \alpha} \, \slashed{n}  \xi_{u}^{\downarrow} ],
\nonumber\\
F_{2}^{\mu}  =&   [ \bar \chi_{u}^{\downarrow}  \slashed{\bar n} \,
\gamma_{\perp \alpha} \gamma_{\perp \beta}  C^{-1}
\bar \chi_{u}^{\uparrow}  ]  \, [ \bar \chi_{s}^{\uparrow}  \slashed{\bar n} \gamma_{\perp}^{\beta} \gamma_{\perp}^{\alpha}  \,  \gamma_{\perp}^{\mu} \,  \slashed{n}   \xi_{s}^{\uparrow}  ] \, [\xi_{u}^{\uparrow}  C \slashed{n}  \xi_{u}^{\downarrow} ],
\nonumber\\
F_{3}^{\mu}  =&  [ \bar \chi_{u}^{\uparrow}  \slashed{\bar n} \, \gamma_{\perp \rho}  C^{-1}  \bar \chi_{u}^{\uparrow}  ]  \, [ \bar \chi_{s}^{\downarrow}  \slashed{\bar n} \, \gamma_{\perp}^{\rho} \,  \gamma_{\perp}^{\mu} \, \gamma_{\perp}^{\tau} \gamma_{\perp}^{\alpha} \gamma_{\perp}^{\beta} \, \slashed{n}   \xi_{s}^{\downarrow}  ] \, [ \xi_{u}^{\uparrow}  C \gamma_{\perp \tau}   \gamma_{\perp \beta} \gamma_{\perp \alpha} \, \slashed{n}  \xi_{u}^{\uparrow} ],
\nonumber \\
F_{4}^{\mu}  =& [ \bar \chi_{u}^{\uparrow}  \slashed{\bar n} \, \gamma_{\perp \alpha} \gamma_{\perp \beta} \gamma_{\perp \rho}  C^{-1}  \bar \chi_{u}^{\uparrow}  ]  \, [ \bar \chi_{s}^{\downarrow}  \slashed{\bar n} \, \gamma_{\perp}^{\beta}  \gamma_{\perp}^{\alpha}  \gamma_{\perp}^{\rho} \,  \gamma_{\perp}^{\mu} \, \gamma_{\perp}^{\tau} \, \slashed{n}   \xi_{s}^{\downarrow}  ] \,[  \xi_{u}^{\uparrow}  C \gamma_{\perp \tau}   \, \slashed{n}  \xi_{u}^{\uparrow} ] \,,
\label{eq:operator basis}
\end{align}
where we have introduce addtional seven so-called evanescent operators.
The $\xi$ represents the collinear quark field propagating in the $\bar n$ direction, while the $\chi$ corresponds to the anti-collinear quark field associated with the $n$ direction.
The evanescent operators $E_i^{\mu}$ and $F_i^{\mu}$ have distinct origins.
Specifically, the operators $E_i^{\mu}$ originate from loop corrections to the correlation function $\Pi_\mu$, while the operators $F_i^{\mu}$ arise from loop corrections to the physical operators.
By reducing the dimension to four dimensions and applying the Fierz transformation, we can verify that the seven evanescent operators $E_i^{\mu}$ and $F_i^{\mu}$ vanish.

By matching the matrix element $\Pi_\mu$ onto the collinear operator matrix elements, one arrives at
\begin{align}
 \frac{(4 \pi \alpha_s)^2}{Q^6} \, \sum_{k}  \,
T_{k} \otimes \langle {\cal O}_{k}^{\mu} \rangle
=\frac{(4 \pi)^4}{Q^6} \, \sum_{k}  \,  \sum_{\ell}  \,
\left ( {Z_{\alpha} \alpha_s \over  4 \pi} \right )^{\ell + 2} \, A_{k}^{(\ell)} \,
\otimes \langle {\cal O}_{k}^{\mu} \rangle^{(0)} \,.
\label{matching conditions}
\end{align}
The $T_k$ are the UV renormalized and IR subtracted hard coefficients, which are finite.
We have used the fact that the loop corrections to the collinear operator matrix elements are scaleless when evaluated between collinear external states. Therefore, the collinear operator matrix elements are reduced to their tree-level matrix elements on the right-hand side of \eqref{matching conditions}.
The amplitudes $A_{k}^{(\ell)}$ denote the bare $\ell$-loop QCD contributions, containing $1/\epsilon$ poles originating from UV and/or IR divergences.
On the other hand, the effective matrix elements of the renormalized collinear operators take the form
\begin{eqnarray}
\langle {\cal O}_{k}^{\mu} \rangle = \sum_{m} \, \sum_{\ell}  \, \left ( {\alpha_s \over 4 \pi} \right )^{\ell} \,
Z_{k m}^{(\ell)} \otimes  \langle {\cal O}_{m}^{\mu} \rangle^{(0)}  \,,
\end{eqnarray}
where $Z_{k m}^{(\ell)}$ stands for the ${\ell}$-loop matrix kernel of renormalization constants.
By further expanding the short-distance functions $T_k$ in the matching relation~(\ref{corr}) in powers of $\alpha_s$, we are able to derive the master formulae for the tree-level and one-loop short-distance coefficients associated with the two physical basis defined in~\eqref{eq:operator basis}
\begin{eqnarray}
T_{k}^{(0)} &=&  A_{k}^{(0)} \,,
\nonumber \\
T_{k}^{(1)} &=&  A_{k}^{(1)} + 2 \,  Z_{\alpha}^{(1)} \, A_{k}^{(0)}
- \sum_{k} \, Z_{m k}^{(1)}  \otimes T_{m}^{(0)}.
\label{master formula}
\end{eqnarray}  
Following the prescription established in~\cite{Buras:1989xd,Buras:1992tc,Dugan:1990df,Herrlich:1994kh,Beneke:2005vv}, the renormalization constants associated with the evanescent operators are fixed by requiring that the infrared-finite matrix elements $\langle E_i^{\mu} \rangle$ vanish identically in four dimensions. 
The renormalization constants $Z_{\Sigma\Sigma}^{(1)}$ and $Z_{TT}^{(1)}$ have been derived in~\cite{Braun:1999te,Huang:2025bbk}.
To determine the evanescent-to-physical mixing kernels explicitly, we can start by considering the $\alpha_s$ expansion of the renormalized matrix elements of operators
\begin{align}
    \langle {\mathcal{O}}_{a}^\mu\rangle=&\left\{\delta_{ab}+\dfrac{\alpha_s}{4\pi}\left[M_{ab}^{(1)}+Z_{ab}^{(1)}\right]+\mathcal{O}(\alpha_s^2)\right\}\otimes \langle {\mathcal{O}}_{b}^\mu\rangle^{(0)},\label{eq:operator renormalizeOff}
\end{align}
The matrix elements $M_{ij}^{(1)}$ are obtained with dimensional regularization applied only to UV divergences but with the IR regularization scheme being different from the dimensional one. Renormalizing the matrix elements of the evanescent operators to zero yields the
relations
\begin{align}
    Z_{E_k \Sigma}^{(1)}=-M^{(1)}_{E_k \Sigma},\quad Z_{E_k T}^{(1)}=-M^{(1)}_{E_k T}\, .
\end{align}
Among them, we find $M_{E_2 \Sigma}^{(1)} = M_{E_3 T}^{(1)} = 0$ and $ M_{E_{1,2} T}^{(1)} = \mathcal{O}(\epsilon)$, resulting in $Z_{E_2 \Sigma}^{(1)}=Z_{E_k T}^{(1)}=0$.
As a result, the nontrivial mixing between evanescent and physical operators induces non-vanishing {finite} renormalization constants  $Z_{E_i\Sigma}^{(1)}$. 
These finite terms are essential for consistently establishing the perturbative factorization formula within the $D$-dimensional framework.

\section{Scheme dependence of short-distance coefficients}
It can be inferred from~(\ref{hard-collinear factorization formula}) that the short-distance coefficient is sensitive only to the evanescent scheme $F_i$, while it remains independent of $E_i$, because the hyperon distribution amplitude is only related to $F_i$.

To show the dependence of the short-distance coefficient on the evanescent scheme, we can apply the following transformation to the operators
\begin{align}
    & \mathcal{O}_{\Sigma}^{\prime \mu} = \mathcal{O}_{\Sigma}^{\mu} \,,  \quad E_i^{\prime \mu} = E_i^\mu - \epsilon \, a_i \mathcal{O}_{\Sigma}^\mu \,. 
\end{align}
Without the loss of generality, we have set $\mathcal{O}_T^\mu$ to zero, for simplicity (so as $\mathcal{O}_T^{\prime \mu}$).

Under this transformation, we assume that the transformed $T_\Sigma^{(1)}$ is $T_\Sigma^{\prime (1)}$.
From the \eqref{master formula}, we find that we can express the following transformed quantities with the original ones to obtain the relation between $T_\Sigma^{\prime (1)}$ and $T_\Sigma^{(1)}$
\begin{align}
    A_\Sigma^{\prime (0)} \,,A_\Sigma^{\prime (1)} \,, A_{E_i}^{\prime (0)} \,, Z_{\Sigma\Sigma}^{\prime (1)} \,, Z_{E_i\Sigma}^{\prime (1)} \,.
    \label{eq:needAZ}
\end{align}
Later we will see that since the physical operator does not mix into  $E_i$ ($Z_{\Sigma E_i} \equiv 0$), the short-distance coefficient will not depend on the choice of $a_i$.
Since the QCD amplitudes are scheme-independent, we have the following relation
\begin{align}
        & A_k^{\prime (\ell)} \otimes \langle \mathcal{O}_k^{\prime \mu} \rangle^{(0)} = A_k^{(\ell)} \otimes \langle \mathcal{O}_k^\mu \rangle^{(0)} \,,
        \nonumber\\
        & \mathrm{LHS} = A_\Sigma^{\prime (\ell)} \otimes \langle \mathcal{O}_{\Sigma}^{\prime \mu} \rangle^{(0)} + \sum_i A_{E_i}^{\prime (\ell)} \otimes \langle E_i^{\prime \mu} \rangle^{(0)} \,,
        \nonumber\\
        & \mathrm{RHS} = A_\Sigma^{(\ell)} \otimes \langle \mathcal{O}_\Sigma^{\mu} \rangle^{(0)} + \sum_i A_{E_i}^{(\ell)} \otimes \langle E_i^{\mu} \rangle^{(0)} \,,
        \label{eq:QCDAmp}
    \end{align}
 Comparing the coefficient of $\langle E_i^{\mu} \rangle^{(0)}$ and $\langle \mathcal{O}_\Sigma^\mu \rangle^{(0)}$ on the both side of \eqref{eq:QCDAmp}, we obtain
\begin{align}
    A_{E_i}^{\prime (0)} =& A_{E_i}^{(0)} \,,
\nonumber\\
    A_\Sigma^{\prime (\ell)} =& 
    A_\Sigma^{(\ell)} + \sum_i \epsilon \, a_i A_{E_i}^{(\ell)} \,.%
    \label{ANbox}
\end{align}
Since the matrix elements of the bare operators preserve the linear relations of the basis transformations $\langle \mathcal{O}_k^{\prime \mu} \rangle_{\mathrm{bare}}^{(1)} = \langle \mathcal{O}_k^\mu \rangle_{\mathrm{bare}}^{(1)}$, we can obtain
\begin{align}
    Z_{\Sigma\Sigma}^{\prime (1)} =& Z_{\Sigma\Sigma}^{(1)} \,,
\nonumber\\
    Z_{E_i\Sigma}^{\prime (1)} =& Z_{E_i\Sigma}^{(1)} - \epsilon \, a_i Z_{\Sigma\Sigma}^{(1)} + \sum_j \epsilon \, a_j Z_{E_i E_j}^{(1)} \,.
\end{align}
Till now, we have found all the transformation rules listed in \eqref{eq:needAZ}. Collecting all the pieces together, we arrive at the scheme dependence of the short-distance coefficient
\begin{align}
    T_{\Sigma}^{\prime (0)} =& T_{\Sigma}^{(0)} + \epsilon \sum_i a_i T_{E_i}^{(0)} \,, \quad T_{E_i}^{\prime (0)} = T_{E_i}^{(0)} \,,
\nonumber\\
    T_{\Sigma}^{\prime (1)} =& T_{\Sigma}^{(1)} + \epsilon \sum_i a_i T_{E_i}^{(1)}\,.
\end{align}
Since $T_{E_i}^{(\ell)}$ are finite, one can immediately conclude that the physical short-distance coefficient is free of the evanescent scheme $E_i$ in the limit $\epsilon \to 0$.
The demonstration of the scheme dependence associated with $F_i$ follows along the same lines and will therefore not be presented here.

%
\section{Next-to-Leading-Order QCD Computations}
%
In this section, we describe the one-loop QCD computation of the bare amplitude $\Pi_\mu$.
The NLO Feynman diagrams are generated using \texttt{FeynArts}~\cite{Hahn:2000kx}. There are 48 diagrams generated at tree-level, and 2040 diagrams generated at one-loop level.
%

\begin{figure}[h]
    \centering
    \includegraphics[width=0.6\linewidth]{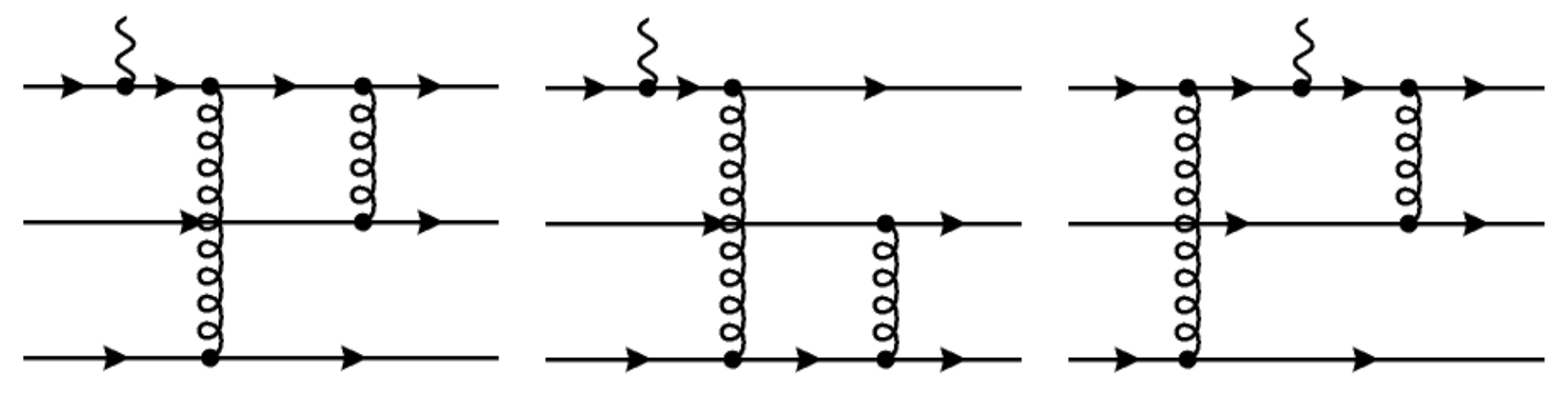}
    \caption{Independent tree-level diagrams contributing to the QCD amplitude $\Pi_\mu$.}
    \label{fig:tree-diagrams}
\end{figure}

The three independent tree-level diagrams contributing to the QCD amplitude $\Pi_\mu$ are shown in FIG~\ref{fig:tree-diagrams}. 

To compute the $\Pi_\mu$, the Passarino-Veltman reduction~\cite{Passarino:1978jh} to the tensor structure of the amplitude is implemented first.
The yielding one-loop scalar integrals are further reduced to a small set of master integrals by IBP relations with the help of {\tt FIRE}~\cite{Smirnov:2008iw}.
Exploiting further the fact that the momenta of external quark states are either in collinear or anti-collinear direction, we arrive at two independent master integrals at one loop.
The above workflow, including the identification of equivalent integrals under variable transformations, has been implemented in the package \texttt{LoopS}~\cite{shi_2025_17383900}.

Inserting the conformal expansion of the twist-three hyperon distribution amplitude~\eqref{eq:conformal expansion}
into the one-loop factorization formula for the Dirac hyperon form factor
and performing the resulting four-fold integrals over the momentum fractions analytically, we obtain
\begin{align}
Q^4 \, F_1^{\Sigma^+}(Q^2) =&  {(4  \pi  \alpha_s)^2} \, 
{100 \over 3} \,  f_\Sigma ^2  \,  \sum_{m, \, n, \, m^{\prime}, \,  n^{\prime}} \,
 \varphi_{m n}  \,
 \left [ e_u \,  {\cal {T}}_{m n}^{m^{\prime} n^{\prime}}
+ e_s \, {\cal {\widetilde{T}}}_{m n}^{m^{\prime} n^{\prime}} \right ] \varphi_{m^{\prime} n^{\prime}}  \,,
\end{align}
The $\alpha_s$ expansions of $\mathcal{T}_{mn}^{m'n'}$ and $\widetilde{\mathcal{T}}_{mn}^{m'n'}$ up to NLO are
\begin{align}
{\cal T}_{m n}^{m^{\prime} n^{\prime}} =&  {\cal N}_{m n}^{m^{\prime} n^{\prime}} \,
\left \{  1 +  {\alpha_s  \over 4 \, \pi}  \,
\left [  2 \, \beta_0 \, L_R + C_F \, \gamma_{mn}^{m'n'}   \,
L_F  + {\cal R}_{m n}^{m^{\prime} n^{\prime}} \right ]
\right \}  \,,
\nonumber \\
{\cal \widetilde{T}}_{m n}^{m^{\prime} n^{\prime}} =&  {\cal \widetilde{N}}_{m n}^{m^{\prime} n^{\prime}} \,
\left \{  1 +  {\alpha_s  \over 4 \, \pi}  \,
\left [  2 \, \beta_0 \, L_R + C_F \, \gamma_{mn}^{m'n'}   \,
L_F  + {\cal \widetilde{R}}_{m n}^{m^{\prime} n^{\prime}} \right ]
\right \}  \,,
\end{align}
where $L_R = \ln{(\nu^2 / Q^2)}$ and $L_F = \ln{(\mu_F^2 / Q^2)}$. $\nu$ is the renormalization scale.
Although the full integrand is finite, individual terms may develop spurious singularities
either inside the integration domain or at its boundaries.
To deal with this issue, we first rewrite the result in terms of Goncharov polylogarithms~\cite{Goncharov:1998kja}.
Spurious singularities located inside the integration region are straightforward to handle,
since the integrand remains analytic.
In this case, we evaluate the primitive functions analytically with the aid of
{\tt PolyLogTools} and obtain the correct result by substituting the integration limits.
By contrast, spurious singularities at the integration boundaries require a regulator.
We introduce such a regulator by truncating the integration range and subsequently
expanding in the truncation parameter to extract the finite result.

We have verified that the coefficients $\gamma_{mn}^{m'n'}$ coincide with the sum of the one-loop anomalous dimensions of the local moments $f_\Sigma\,\varphi_{m^{(\prime)} k^{(\prime)}}$ of the twist-three hyperon distribution amplitude $\varphi_\Sigma$
\begin{align}
    \gamma_{mn}^{m'n'} = \gamma_{m n}^{(1)} + \gamma_{m^{\prime} n^{\prime}}^{(1)} \,.
\end{align}
The tree-level coefficients ${\cal N}$ and ${\cal \widetilde{N}}$, as well as the NLO kernels ${\cal R}$ and ${\cal \widetilde{R}}$, are turned out to be identical with those in~\cite{Huang:2024ugd}. For the convenience of the reader, we provide the numerical values of the first few elements of 
${\cal N}$, ${\cal \widetilde{N}}$, ${\cal R}$, and ${\cal \widetilde{R}}$ in TABLE.~\ref{tab:NRvalues}.

\begin{table}[h]
\centering
\caption{Numerical values of the first few elements of 
${\cal N}$, ${\cal \widetilde{N}}$, ${\cal R}$, and ${\cal \widetilde{R}}$.}
\label{tab:NRvalues}
\begin{tabular}{|c|cc|cc|}
\hline
$(mn,m'n')$ 
& $\mathcal{N}_{mn}^{m'n'}$ 
& $\widetilde{\mathcal{N}}_{mn}^{m'n'}$ 
& $\mathcal{R}_{mn}^{m'n'}$ 
& $\widetilde{\mathcal{R}}_{mn}^{m'n'}$ \\ \hline
(00,00) & 1.00  & 2.00   & 87.34  & 71.96 \\ \hline
(10,00) & 8.17  & -8.17  & 99.25  & 99.25 \\
(10,10) & 168.78& 174.22 & 116.34 & 115.15 \\ \hline
(11,00) & 3.50  & -3.50  & 115.87 & 74.00 \\
(11,10) & 38.11 & -38.11 & 118.94 & 120.65 \\
(11,11) & 59.89 & 21.78  & 108.59 & 105.28 \\ \hline
(20,00) & 21.35 & 8.05   & 106.88 & 99.74 \\
(20,10) & 89.83 & -89.83 & 126.80 & 127.58 \\
(20,11) & 93.10 & 9.80   & 129.65 & 153.27 \\
(20,20) & 234.71& 54.88  & 139.22 & 136.02 \\ \hline
(21,00) & 8.75  & -8.75  & 110.76 & 112.62 \\
(21,10) & 159.25& 159.25 & 132.17 & 133.58 \\
(21,11) & 44.92 & -44.92 & 130.09 & 132.15 \\
(21,20) & 84.53 & -84.53 & 143.23 & 145.55 \\
(21,21) & 159.25& 159.25 & 151.03 & 152.11 \\ \hline
(22,00) & 3.20  & 0.10   & 111.71 & 194.80 \\
(22,10) & 12.25 & -12.25 & 128.59 & 128.85 \\
(22,11) & 7.12  & 2.68   & 159.72 & 142.92 \\
(22,20) & 30.35 & 4.94   & 146.17 & 150.39 \\
(22,21) & 7.18  & -7.18  & 155.30 & 158.03 \\
(22,22) & 7.79  & 1.72   & 141.40 & 141.08 \\
\hline
\end{tabular}
\end{table}

It is straightforward to check that the NLO QCD result for the Dirac hyperon form factor obtained from~\eqref{hard-collinear factorization formula} is indeed independent of the factorization scale $\mu_F$ at one-loop order.
This cancellation follows from the one-loop RG evolution equation of the leading-twist hyperon distribution amplitude $\varphi_\Sigma$~\cite{Braun:1999te, Huang:2025bbk}.
To achieve the resummation of large logarithms $\ln(Q^2/\Lambda_{\rm QCD}^2)$ at NLL accuracy in the factorization formula~\eqref{hard-collinear factorization formula}, the complete two-loop RG evolution of the hyperon distribution amplitude $\varphi_\Sigma$ is required.
Such a two-loop evolution equation has been determined very recently, both in our evanescent scheme defined in~\eqref{eq:operator basis} and in the KM scheme~\cite{Huang:2025bbk}.

\section{Numerical Analysis}
In this section, we investigate the phenomenological implications of the NLO QCD corrections to the Dirac hyperon form factor.
To this end, we first specify the essential nonperturbative input, namely the twist-three hyperon distribution amplitude, which enters the hard-collinear factorization formula for the Dirac hyperon form factor.
\begin{figure}[h]
    \centering
    \includegraphics[width=0.5\linewidth]{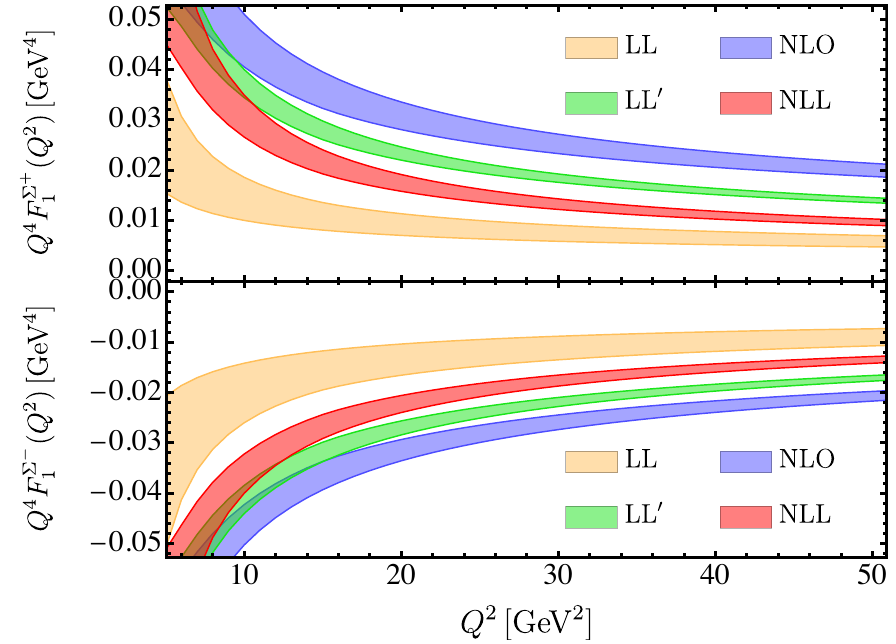}
    \caption{Predictions for the leading power hard-gluon-exchange contributions to the Dirac form factors of the $\Sigma^+$ (upper panel) and $\Sigma^-$ (lower panel) at LL, NLO, LL$^\prime$, and NLL accuracy, based on the LAT25 hyperon distribution amplitude. The colored bands represent perturbative uncertainties from scale variations $\nu^2 = \mu_F^2 = \langle x \rangle Q^2$ with $1/6 \le \langle x \rangle \le 1/2$.}
    \label{fig:pheno}
\end{figure}
For illustrative purposes, we adopt the recent lattice QCD results~\cite{Bali:2024oxg}
\begin{align}
    &\{f_\Sigma^{\rm KM} (\mu_0), \varphi_{10}^{\rm KM} (\mu_0), \varphi_{11}^{\rm KM} (\mu_0), \varphi_{20}^{\rm KM} (\mu_0), \varphi_{21}^{\rm KM} (\mu_0), \varphi_{22}^{\rm KM} (\mu_0)\} 
\nonumber\\
    =& \{5.32, 0.094, 0.200, 0, 0, 0\} \times 10^{-3} \, {\rm GeV}^2 \,,
\end{align}
where reference scale $\mu_0 = 2 \, {\rm GeV}$. The nonperturbative quantities labeled with the superscript ``KM'' are defined in the KM scheme.
For the purpose of numerical analysis within our framework, these moments must be converted to our evanescent scheme.
Applying the two-loop conversion factors derived in~\cite{Huang:2025bbk}, we obtain
\begin{align}
    &\{f_\Sigma (\mu_0), \varphi_{10} (\mu_0), \varphi_{11} (\mu_0),
      \varphi_{20} (\mu_0), \varphi_{21} (\mu_0), \varphi_{22} (\mu_0)\}
\nonumber\\
    =& \{5.13, 0.092, 0.197, -0.002, 0.000, 0.000\}
    \times 10^{-3} \, \mathrm{GeV}^2 \,.
\end{align}
In FIG.~\ref{fig:pheno}, we present the resulting theoretical predictions for these fundamental hadronic observables obtained with the \texttt{LAT25} model at LL, NLO, LL$^\prime$, and NLL accuracy.
The LL$^\prime$ approximation is defined by incorporating the fixed-order NLO correction into the LL-resummed result.
It is observed that, within the kinematic range $Q^2 \in [20,\,50]\,\mathrm{GeV}^2$, the one-loop corrections to the short-distance coefficient of the hard-gluon-exchange contribution remain significant relative to the tree-level result, although their relative size decreases as $Q^2$ increases. Notably, upon including NLL resummation effects, the magnitude of the one-loop corrections in the interval $Q^2 \in [20,\,50]\,\mathrm{GeV}^2$ is reduced by approximately $20\%$--$30\%$ compared to the LL$^\prime$ approximation.

\section{Conclusions}
In this work, we have performed a systematic next-to-leading-order QCD analysis of the leading power hard-gluon-exchange contributions to the Dirac electromagnetic form factors of the $\Sigma$ hyperons within the hard-collinear factorization formalism.
Adopting a consistent treatment of evanescent operators, the NLO short-distance coefficient function $H_\Sigma$ has been extracted analytically from the relevant seven-point partonic amplitudes at ${\cal O}(\alpha_s^3)$, with UV renormalization and IR subtractions implemented in a rigorous way.
The resulting coefficient is finite and exhibits an explicit dependence on the evanescent scheme.
Combining this result with the conformal expansion of the twist-three $\Sigma$ distribution amplitude, we have explored the numerical impact of NLO corrections and NLL resummation effects using representative nonperturbative input.

%
\begin{acknowledgments}
\section*{Acknowledgements}

This research is supported by the National Natural Science Foundation of China with Grants No. 12475097 and No.12535006, and by the Natural Science Foundation of Tianjin with Grant No. 25JCZDJC01190. Xue-Chen Zhao is supported by the Postdoctoral Fellowship Program of CPSF under Grant Number GZB20250794.

\end{acknowledgments}

\bibliographystyle{apsrev4-1}
\bibliography{References}
\end{document}